# Multifrequency superscattering from subwavelength hyperbolic structures

Chao Qian[1,2,⊥], Xiao Lin[3,⊥,*], Yi Yang[4], Fei Gao[3], Yichen Shen[5], Josue Lopez[4], Ido Kaminer[5,*],

Baile Zhang[3,6], Erping Li[1,2], Marin Soljačić[5], and Hongsheng Chen[1,2,*]

[1]*Key Lab. of Advanced Micro/Nano Electronic Devices & Smart Systems of Zhejiang, The Electromagnetics Academy at Zhejiang University, Zhejiang University, Hangzhou 310027, China.*
[2]*ZJU-UIUC Institute, College of Information Science and Electronic Engineering, Zhejiang University, Hangzhou 310027, China.*
[3]*Division of Physics and Applied Physics, School of Physical and Mathematical Sciences, Nanyang Technological University, Singapore 637371, Singapore.*
[4]*Department of Electrical Engineering and Computer Science, Massachusetts Institute of Technology, Cambridge, MA 02139, USA.*
[5]*Department of Physics, Massachusetts Institute of Technology, Cambridge, MA 02139, USA.*
[6]*Centre for Disruptive Photonic Technologies, NTU, Singapore 637371, Singapore.*
[*]*Corresponding Author: xiaolinbnwj@ntu.edu.sg (X. Lin); kaminer@mit.edu (I. Kaminer); hansomchen@zju.edu.cn (H. Chen).*
[⊥]*These two authors contribute equally to this work.*

**Abstract:** Superscattering, i.e., a phenomenon of the scattering cross section from a subwavelength object exceeding the single-channel limit, has important prospects in enhanced sensing/spectroscopy, solar cells, and biomedical imaging. Superscattering can be typically constructed only at a single frequency regime, and depends critically on the inescapable material losses. Under such realistic conditions, superscattering has not been predicted nor observed to exist simultaneously at multiple frequency regimes. Here we introduce multifrequency superscattering in a subwavelength hyperbolic structure, which can be made from artificial metamaterials or from naturally-existing materials, such as hexagonal boron nitride (BN), and show the advantage of such hyperbolic materials for reducing structural complexity. The underlying mechanism is revealed to be the multimode resonances at multiple frequency regimes as appear in BN due to the peculiar dispersion of phonon-polaritons. Importantly, the multifrequency superscattering has a high tolerance to material losses and some structural variations, bringing the concept of multifrequency superscattering closer to useful and realistic conditions.

Scattering of light is ubiquitous in nature. Understanding and especially controlling the scattering effect, dating back to the pioneering work of Rayleigh[1] and Mie[2] a century ago, is of fundamental importance to elementary electromagnetic theory and its many distinct applications.[3-7] With the recent advances in nano-fabrication and the advent of metamaterials[8-10] and two-dimensional (2D) materials,[11-13] the scattering effect in designed structures can be flexibly enhanced or suppressed. To obtain stronger scattering response, typically either the *absolute*[14,15] or the *volume-normalized*[16,17] scattering cross-sections are chosen as the figure of merit, depending on the application scenarios. The superscattering phenomenon[14,15] realizes the larger absolute scattering cross-section (exceeding single-channel limit[14]) by overlapping multiple decoupled resonances in subwavelength objects. Although intensive attentions have been ignited, the phenomenon of superscattering was *mainly* reported to exist at a single frequency regime.[14,15,18-20] The phenomenon of superscattering simultaneously existing at multiple frequency regimes (denoted as multifrequency superscattering below) may enable improved sensitivity and accuracy for many applications, ranging from sensing, optical imaging, optical tagging to spectroscopy.[21-23] Therefore, multifrequency superscattering is still highly desirable but yet to be realized. The continuing pursuit of these extreme scattering phenomena will extend our capability to flexibly control the scattering effect and facilitate more advanced applications.[21-26]

For multifrequency superscattering, multi-layer plasmonic structures, which have multiple polaritonic dispersion lines and thus may support multimode plasmonic resonances simultaneously in multiple frequency regimes, might be the conventional platform to realize this phenomenon. However, many-layer plasmonic structures place challenges for realization, including the detrimental influence of realistic losses on the superscattering performance and the increased complexity of structural design, such as the precise requirement on the thickness of each layer. For example, when considering a hypothetically-small loss, which can be much more severe in realistic plasmonic constituents, the multifrequency superscattering phenomenon in a five-layer plasmonic structure will be deteriorated and thus disappear.[27] Therefore, when considering realistic losses, the realization of the multifrequency superscattering from a subwavelength structure still remains very challenging, even *in theory*. Furthermore, from a practical perspective, it is very desirable to achieve the multifrequency superscattering with reduced structural complexity. A thin hyperbolic structure, from the perspective of polaritonic dispersions, can provide the similar functionality as many-layer plasmonic structures.[11,12,28] This makes the possibility of multifrequency superscattering in hyperbolic structures very attractive.

In this work, we propose to use hyperbolic nanostructures, such as a subwavelength hyperbolic-dielectric two-layer cylindrical rod, to achieve the multifrequency superscattering. The hyperbolic layer can be artificially-made or be a naturally-existing material. Here we adopt a thin slab of hexagonal boron nitride (BN) (which is a natural hyperbolic material within its two reststrahlen bands[11,12,28-32]) as an example. We reveal that the peculiar dispersion of BN's hyperbolic phonon polaritons, which has many dispersion lines within BN's two reststrahlen bands, increases the possibility of multimode resonances at multiple frequency regimes. Therefore, the phenomenon of multifrequency superscattering exists even in the presence of realistic BN's loss. More importantly, multifrequency superscattering is found to have a high tolerance to the general structural variations, including the variation of BN's thickness in the shell and the dielectric permittivity in the core. Due to the combined advantages of the simple two-layer structure and the high tolerance to material losses and structural variations, hyperbolic materials like BN provide a promising versatile platform for demonstrating exotic scattering effects and fostering potential applications.[21-27]

For conceptual clarity, we consider the multifrequency superscattering from a 2D subwavelength rod, which is composed of a hyperbolic-dielectric two-layer structure; see the inset in Figure 1. In this work, such a 2D rod (i.e., with an infinite length) is equivalent to a finite-length 3D rod with its length being much larger than its diameter; see Figure S6. The lossless dielectric in the core region with a radius of $\rho_1$ has a relative permittivity of $\varepsilon_{diel} = 2.1$ at the frequency of interest (such as SiO$_2$).[33] The hyperbolic material in the shell region is exemplified by BN with a thickness of $\rho_2 - \rho_1$, and the BN plane is normal to the $\rho$-direction. Due to the anisotropy of the crystal and the resulting phonon vibrations, BN is a natural type-I hyperbolic material ($\varepsilon_\rho < 0$ and $\varepsilon_\phi > 0$ in the cylindrical coordinate) within the first reststrahlen band (i.e. ~22.3-24.5 THz), and is a type-II hyperbolic material ($\varepsilon_\rho > 0$ and $\varepsilon_\phi < 0$) within the second reststrahlen band (~41.1-48.2 THz).[11,12,28-32] Below experimental data of the bulk layered BN,[31,32] i.e., with the consideration of realistic loss, are used to characterize the permittivity of BN. It shall be emphasized that when the thickness of BN decreases down to monolayer, the optical properties (especially the material loss) of BN can change remarkably.[34] In addition, the length of the B-N bond in the layered BN slab is 1.45 Å (i.e., angstrom),[35,36] while the perimeters of the inner or outer boundaries of our designed structures are both in the scale of millimeter; this way, the latter is at least three order of magnitude larger than the former. Therefore, it shall be reasonable to argue that the properties of BN in the form of the flat slab and in the form of the cylindrical slab are almost the same. Within the two reststrahlen bands, BN can support the propagation of highly-confined hyperbolic phonon polaritons, which have been extensively studied.[11,12,28-32] Under the incidence of a transverse-magnetic (TM, or *p*-polarized) plane wave, the total magnetic field in the surrounding air region can be expressed as[14,37]

$$\bar{H}_{total} = \hat{z} H_0 \sum_{m=-\infty}^{\infty} \left( i^m J_m(k\rho) e^{im\phi} + i^m S_m H_m^{(1)}(k\rho) e^{im\phi} \right) \quad (1)$$

In the above equation, the constant $H_0$ is the magnitude of the incident magnetic field; $J_m$ and $H_m^{(1)}$ are the $m^{th}$ order of Bessel function and Hankel function of the first kind, respectively; $k$ is the free space wavevector; the unknown parameter $S_m$ is the scattering coefficient of the $m^{th}$ angular momentum channel and can be analytically obtained (see Supporting Information). With the choice of $H_0 = \sqrt{\frac{Watt}{meter} \frac{\omega \varepsilon_0}{2}}$, the scattering cross section, i.e., the total scattered power over the intensity of the incident wave, can be reduced to $C_{sct} = \sum_{m=-\infty}^{\infty} C_{sct,m}$ with $C_{sct,m} = \frac{2\lambda}{\pi} |S_m|^2$. It is worthy to note that the scattering cross section from a single channel cannot exceed the theoretical single-channel limit, i.e. $C_{sct,m} \leq \frac{2\lambda}{\pi}$ for the 2D case.[14] In order to highlight the scattering cross section from the polaritonic resonance ($m \geq 1$), below we define the total scattering cross section without considering the channel with $m = 0$.

To highlight the underlying physics, we first consider the multifrequency superscattering from an *ideal lossless* structure in Figure 1. For the two-layer hyperbolic-dielectric rod, the inner and outer radiuses are $\rho_1 = 0.12$ μm and $\rho_2 = 2.06$ μm, respectively. The hyperbolic BN is assumed to be lossless. Indeed the designed rod-like structures are just the BN nanotubes. We note that the hollow BN nanotubes with a diameter of up to 200 nm have been experimentally synthesized[38-41] and the phenomenon of multifrequency superscattering can be theoretically realized even from a subwavelength rod with the outmost diameter less than 200 nm; see analysis in the Supporting Information. This way, our designed structure shall be viable in experimental implementation. For such a simple subwavelength structure, the scattering cross section mainly comes from the channels with $m = \pm 1, 2$, as shown in Figure 1a. In particular, for these four channels, the scattering cross sections from each channel almost all reach the value of the single-channel limit at two separate frequencies. These two frequencies are 22.37 THz (the corresponding wavelength in air is 13.41 μm) and 47.54 THz (6.31 μm), respectively. At these two frequencies, the total scattering cross sections are both 3.99 times of the value of the single-channel limit. This indicates that the superscattering phenomenon can emerge simultaneously at two different frequency regimes from such a simple subwavelength structure. In order to gain further insight into the multifrequency superscattering, Figure 1b,c and Figure 1e,f show the field distributions of the scattering field from each channel under a plane wave incidence at 22.37 THz and 47.54 THz, respectively. Figure 1d,g shows the summation of the scattering fields from these two channels. It can be clearly seen that the scattering magnetic fields from channel $|m| = 1$ in Figure 1b,e and from channel $|m| = 2$ in Figure 1c,f are in-phase (out-of-phase) in the forward (backward) direction. Therefore, from Figure 1d, g, it is reasonable to argue that the obtained multifrequency superscattering arises mainly from the constructive interference of scattering fields from different channels in the forward direction.

The underlying mechanism of the multifrequency superscattering above are the multimode resonances of phonon polaritons at multiple frequencies within BN's two reststrahlen bands (as opposed to plasmon polaritons in metallic nanostructures[14,15]); see more in Figure S2. Alternatively, this mechanism can be qualitatively understood from the dispersion of phonon polaritons supported by a thin BN planar slab in Figure 2. In order to achieve the superscattering from a subwavelength rod, multiple highly-confined polariton modes supported by the corresponding planar structure are needed within a certain narrow frequency range. There are generally two ways to realize this. One is to create a relatively flat polaritonic dispersion line within a certain frequency range, such as that in the structure of metal-dielectric-metal.[14] The other is to create multiple polaritonic dispersion lines, such as in a many-layer plasmonic structure.[27] Figure 2 shows that a large number of phonon polariton modes at each frequency within BN's two reststrahlen bands can already be supported by a thin hyperbolic BN slab, without resorting to a more complicated many-layer structural design. This indicates that a thin hyperbolic BN slab, from this perspective, can provide the same functionality as a many-layer plasmonic structure. Therefore, the dispersion of BN's phonon polaritons significantly increases the possibility of the multimode resonance at one or even multiple frequencies by only using a simple two-layer structure. For example, by judiciously designing and optimizing the structure, we show other types of multifrequency superscattering can be realized, such as superscattering simultaneously emerging at more frequency regimes (see discussion below) or multifrequency superscattering contributed from more channels (see Figure S3 and Figure S4). Moreover, the dispersion of BN's phonon polaritons may further enable a high tolerance of the multifrequency superscattering to structural variations (see detailed discussion below).

The performance of multifrequency superscattering in the presence of realistic BN loss is studied in Figure 3a. As can be expected, the realistic loss is detrimental to the performance of the multifrequency superscattering by largely degrading the scattering cross sections from each channel. Besides, due to the loss, the two frequencies with the maximum scattering cross sections are also slightly shifted. Consequently, the total scattering cross section at 22.37 (47.54) THz is decreased to 1.61 (2.19) times the value of the single-channel limit. Importantly, even with realistic losses, at both frequencies, each one of the total scattering cross sections still exceeds the single-channel limit.

The multifrequency superscattering at 22.37 THz and 47.54 THz are visualized in Figure 3b,c, respectively, when considering realistic material losses. The magnetic field distribution is obtained via eq 1, which is consistent with the finite-element simulation (the commercial software COMSOL Multiphysics), where a plane wave is set to be incident from the left side of the subwavelength hyperbolic-dielectric rod. Due to the superscattering, the incident plane-wave field is largely disturbed, and a remarkable "shadow" can be seen behind the rod, which is in accordance with the analysis in Figure 1b-g. The size of the shadow is much larger than the rod diameter. As a comparison, when the material in the hyperbolic shell is replaced with the dielectric in the core in Figure 3d, e, the fields of the incident plane wave are less disturbed by the rod.

In order to facilitate future experimental verifications, Figure 4 shows the high tolerance of multifrequency superscattering to structural variations. Here we vary the thickness of the BN shell in Figure 4a, the permittivity of the dielectric core in Figure 4b, and the material loss in Figure 4c as examples to show the high tolerance of the superscattering phenomenon. The large total scattering cross sections can be maintained in the presence of these structural variations, where the frequencies with the maximum total scattering cross sections are only slightly shifted in Figure 4a,b. This indicates that the phenomenon of superscattering from the simple subwavelength hyperbolic structure has a high tolerance to fabrication imperfections. In addition, although loss is detrimental to the performance of superscattering, Figure 4c shows that the phenomenon of superscattering still exists even when the loss of BN increases artificially up to four times. Indeed the loss in BN can be largely suppressed through the isotopic enrichment of BN in experiments.[42-44] It is worthy to note that BN is not the only hyperbolic material; there is an abundant number of choices for hyperbolic materials[8-10] at many different spectral regimes. Therefore, by replacing BN with a hyperbolic metamaterial at the microwave regime (which offers the convenience of macroscopic operation) the simple structure we propose can provide insightful guidance for the future experimental demonstration of the multifrequency superscattering.

Further optimization can enable superscattering at more than two frequency regimes, as shown in Figure 5. For the subwavelength two-layer rod in Figure 5, the inner and outer radiuses are $\rho_1 = 0.01$ μm and $\rho_2 = 3.31$ μm, respectively. Under the ideal lossless assumption, Figure 5a shows that superscattering can simultaneously appear at five different frequency regimes. All these frequency regimes are within BN's first reststrahlen band. Even in the presence of BN's loss, the superscattering can still be maintained at three different frequency regimes.

In conclusion, we propose that a subwavelength hyperbolic-dielectric two-layer structure can be a promising and versatile platform for the demonstration of multifrequency superscattering phenomena. The hyperbolic material can be artificially-made like metamaterials or naturally-existing like BN. Due to the multimode resonance of phonon polaritons at multiple frequencies within BN's two reststrahlen bands, the multifrequency superscattering is found to have a high tolerance to the presence of realistic material losses and small structural variations. This thus can provide a simple yet viable way for future experimental demonstration, and may enable many infrared applications such as chemical and medical sensing.[24,25] In addition, due to the peculiar dispersion of polaritons in a hyperbolic (BN) slab, more exotic scattering phenomena are expected in such a platform, such as simultaneous multifrequency superscattering and multifrequency cloaking.[45,46]

**Acknowledgement**
This work was sponsored by the National Natural Science Foundation of China under Grants No. 61625502, No. 61574127, No. 61601408, No. 61775193 and No. 11704332, the ZJNSF under Grant No. LY17F010008, the Top-Notch Young Talents Program of China, the Fundamental Research Funds for the Central Universities, the Innovation Joint Research Center for Cyber-Physical-Society System, the Nanyang Technological University for Nanyang Assistant Professorship Start-Up Grant, the Singapore Ministry of Education under Grants No. MOE2015-T2-1-070 and No. MOE2011-T3-1-005, and Tier 1 RG174/16(S). Y.Y. was partly supported by the MRSEC Program of the National Science Foundation under Grant No. DMR-1419807. J. J. López is supported in part by a NSF Graduate Research Fellowship under award No. 1122374 and by a MRSEC Program of the National Science Foundation under award No. DMR-1419807.

**Author contributions**
X. L., C. Q. and H. C. initiated the idea. C. Q. performed the main calculation. C. Q., X. L., Y. Y., F. G., Y. S., J. L., I. K., B. Z., E. L., M. S. and H. C. analyzed data, interpreted detailed results and contributed extensively to the writing of the manuscript. X. L., I. K., B. Z., M. S. and H. C. supervised the project.

**Competing financial interests**
The authors declare no competing financial interests.

**Supporting information**
This part includes more discussion of the structural setup, the properties of BN's phonon polaritons, the detailed calculation of scattering cross section, and the phenomenon of multifrequency superscattering.

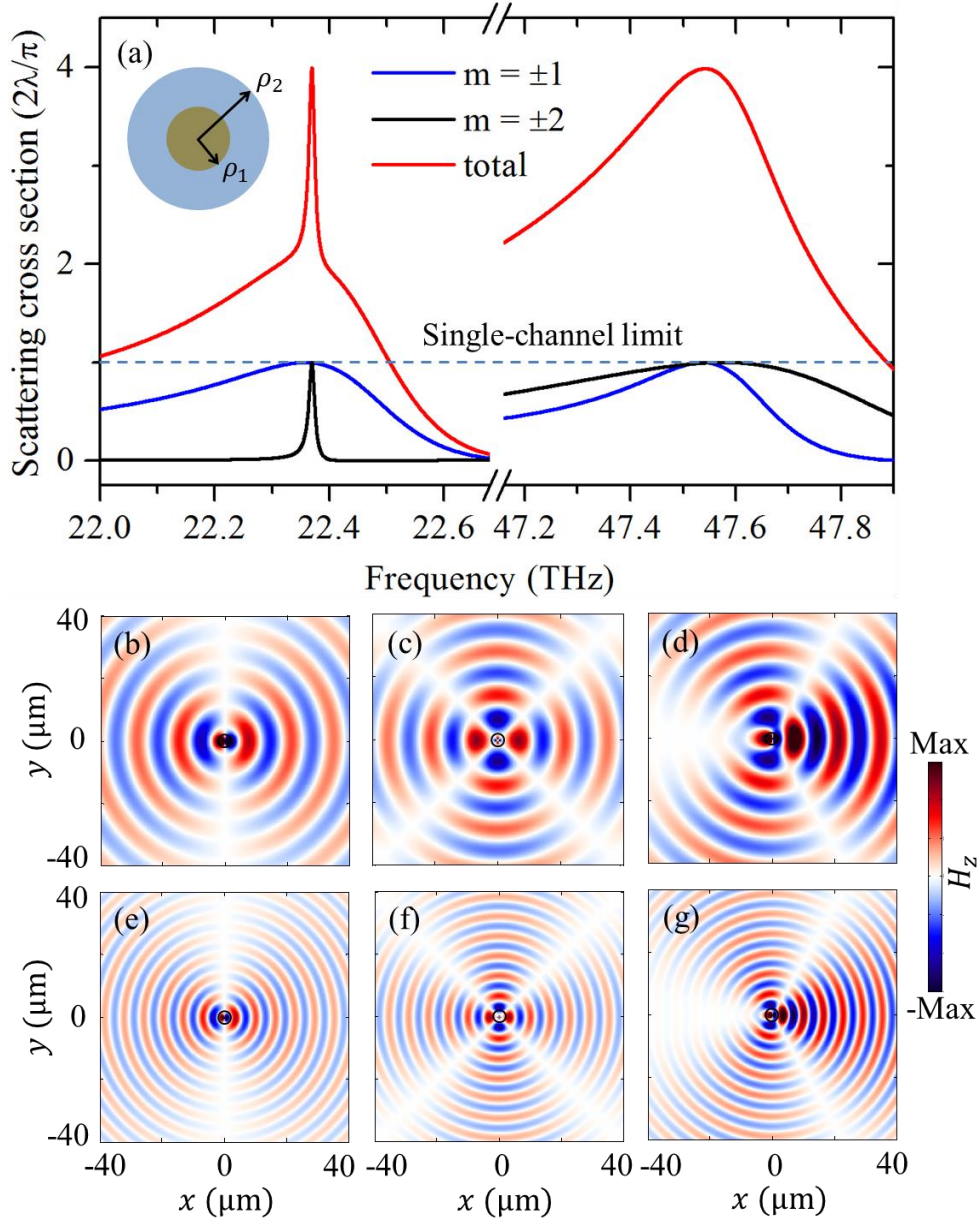

**Figure 1**. Multifrequency superscattering from a subwavelength rod under the ideal lossless assumption. (a) Scattering cross section. The subwavelength two-layer rod is composed of a dielectric core with a relative permittivity of 2.1 (brown region) and a hyperbolic BN shell (blue region); see the inset. The blue dashed line represents the single-channel limit of the scattering cross section. (b-g) Scattering magnetic field distribution at (b-d) 22.37 THz and (e-g) 47.54 THz under a TM plane wave incidence from the left side. The scattering field comes from (b,e) channel $|m| = 1$ and (c,f) channel $|m| = 2$, and (d,g) show their summation. The rod is denoted as a circle in (b-g).

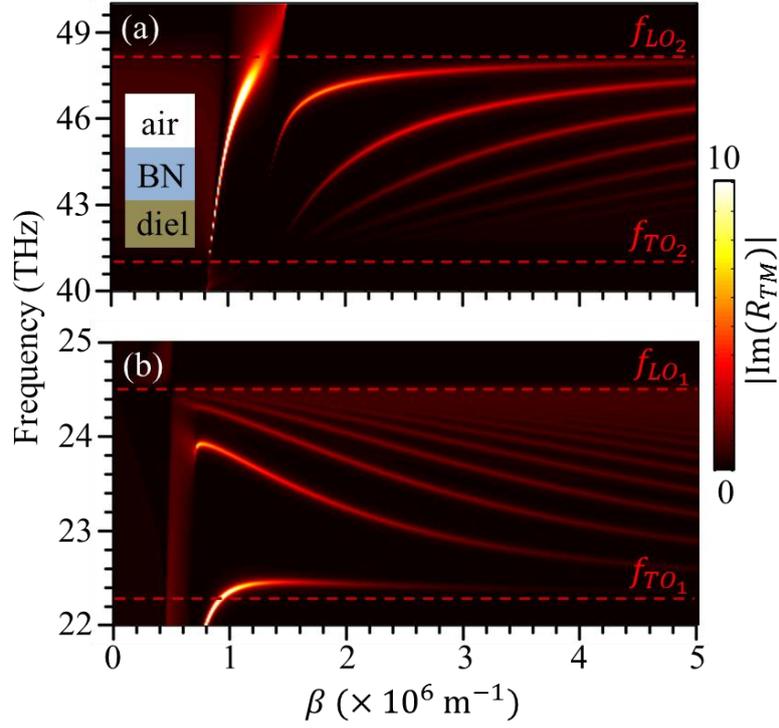

**Figure 2**. Dispersion of phonon polaritons within BN's (a) second and (b) first reststrahlen bands via a false-color plot of $|\text{Im}(R_{TM})|$. $R_{TM}$ is the reflection coefficient of TM plane wave for the planar air-BN-dielectric structure; see the inset in (a). The dielectric is abbreviated as "diel" in (a). The red dashed lines correspond to the transverse optical ($f_{TO}$) and longitudinal optical ($f_{LO}$) frequencies, respectively. Other parameters are the same as those in Figure 1, including the thickness of BN being 1.94 μm and the relative permittivity of the dielectric being 2.1.

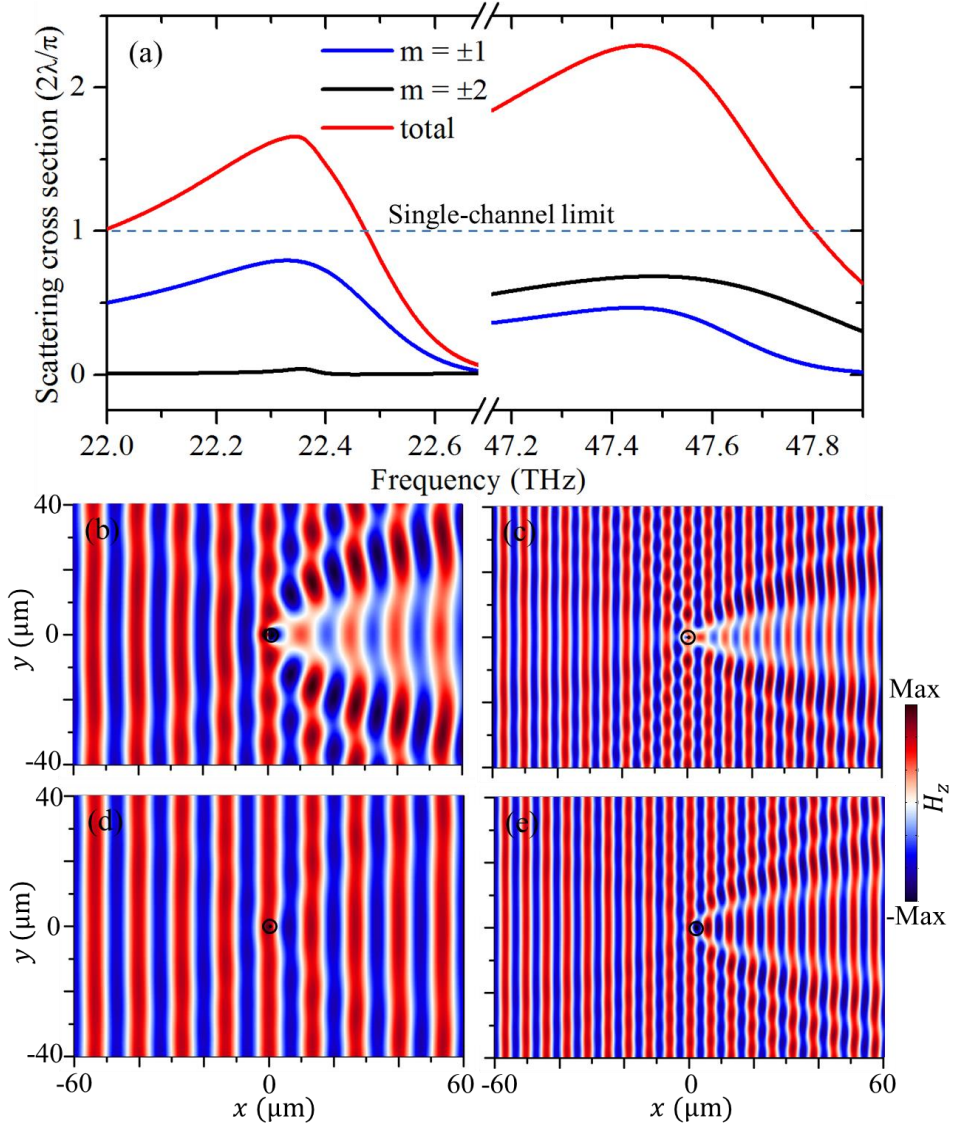

**Figure 3**. Multifrequency superscattering from a subwavelength rod in the presence of material losses. The structure in (a-e) is the same as that in Figure 1, except that for the shell region the hyperbolic BN is replaced with the dielectric in (d-e). The relative permittivity of the core dielectric is 2.1. (a) Scattering cross section. The total scattering cross sections both at 22.37 THz and 47.54 THz are larger than the single-channel limit (dashed line). (b-e) Total magnetic field distribution at (b,d) 22.37 THz and (c,e) 47.54 THz under a TM plane wave incidence from the left side. The rod is denoted as a circle in (b-e).

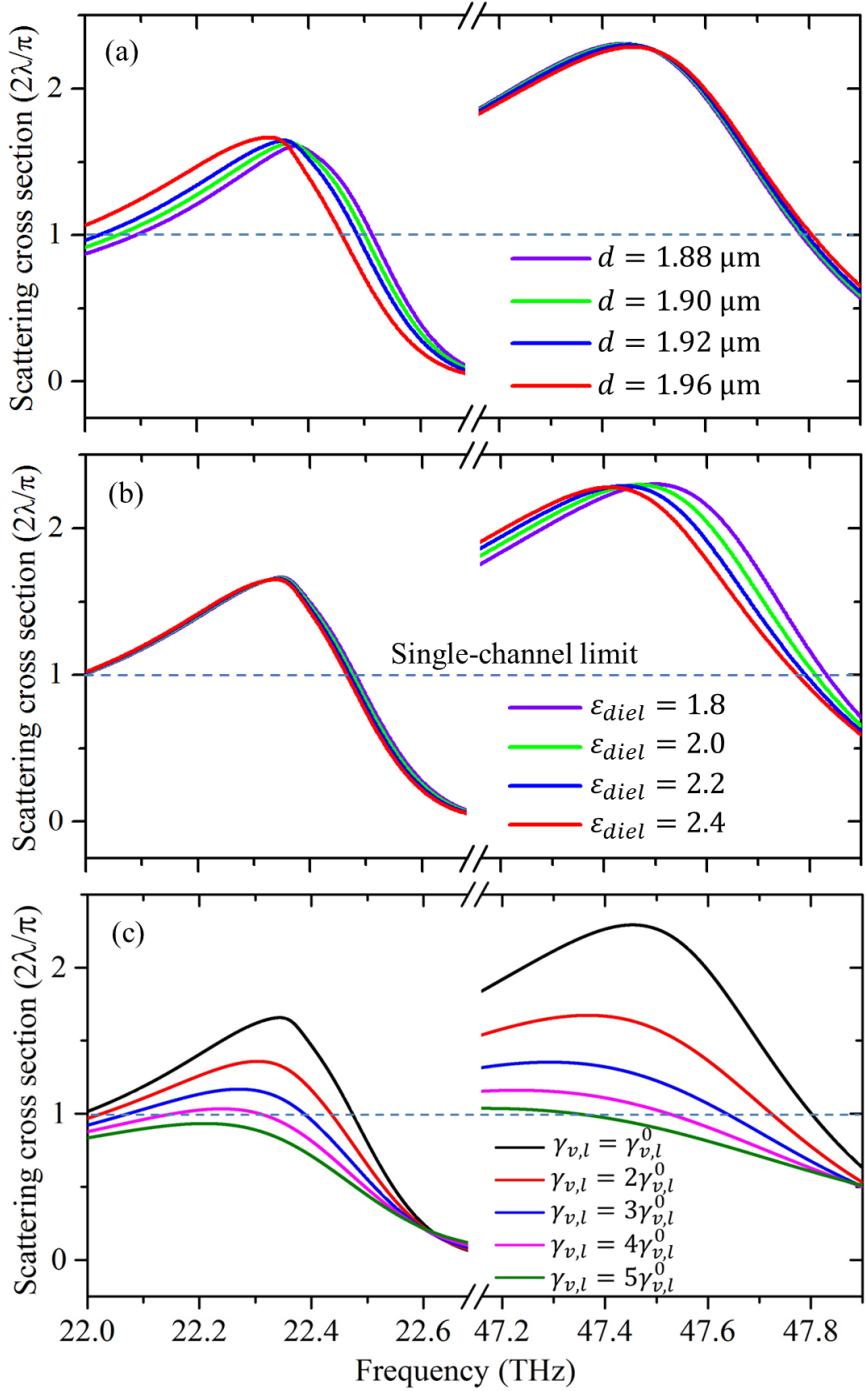

**Figure 4**. The influence of structural variations on the performance of multifrequency superscattering in the presence of BN's loss. The structural setup is the same as that in Figure 1, except (a) the thickness of BN (denoted as $d$), (b) the relative permittivity of the dielectric (denoted as $\varepsilon_{diel}$) or (c) the material loss. The magnitude scattering rate $\gamma_{v,l}$ ($l = \perp, \parallel$) characterizes the loss in BN; a larger value of $\gamma_{v,l}$ indicates a larger loss in BN. $\gamma_{v,l}^0$ is adopted from an experimental data for the bulk layered BN.[31,32]

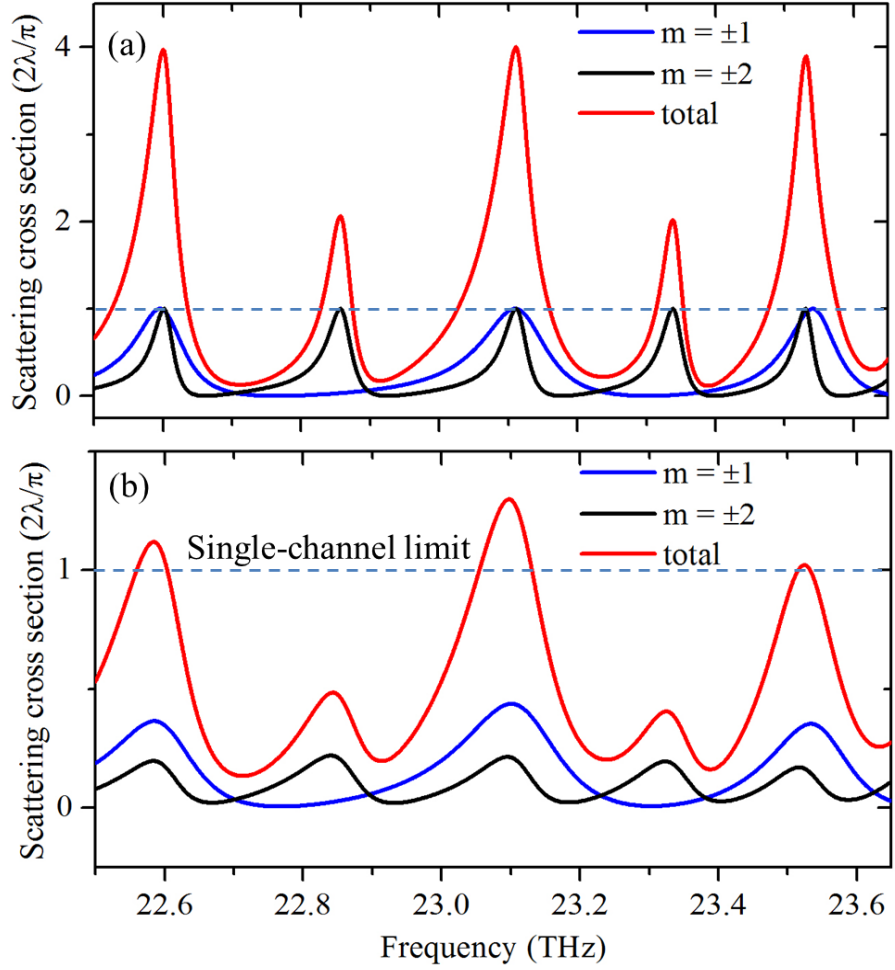

**Figure 5.** Multifrequency superscattering from a subwavelength rod. Scattering cross section (a) under the ideal lossless assumption and (b) in the presence of material losses. The structure is the same as that in Figure 1, except that $\rho_1 = 0.01$ μm and $\rho_2 = 3.31$ μm. The relative permittivity of the core dielectric is 2.1.

For Table of Contents Use Only

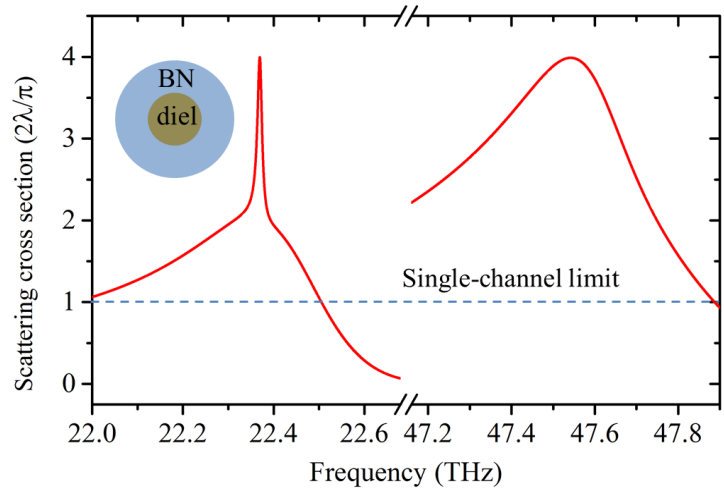

Description: This graph shows the multifrequency superscattering, i.e. with the scattering cross section exceeding the single channel limit at multiple frequency regimes, from a subwavelength hyperbolic structure. The hyperbolic structure can be made from artificial metamaterials or from naturally-existing materials, such as hexagonal boron nitride (BN).